\documentstyle[12pt,epsf]{article}

\def\dsla{ \partial\hspace{-6pt}/ }
\def\psla{ p\hspace{-5pt}/ }

\def\Tr{{\rm Tr}}
\def\str{{\rm str}}

\def\Asla{ A\hspace{-6pt}/ }

\newcommand{\Seff}{S_{\scriptsize\rm eff}}
\newcommand{\Veff}{V_{\scriptsize\rm eff}}

\newcommand{\Leff}{{\cal L}_{\scriptsize\rm eff}}

\newcommand{\LambdaQCD}{\Lambda_{\scriptsize\rm QCD}}

\newcommand{\Sg}[1]{g^{ #1}}

\newcommand{\gammafive}{\gamma _5}

\newcommand{\smp}[1]{p _{#1}}

\begin{document}
\vspace{-1cm}
\noindent
\begin{flushright}
KANAZAWA-2000-01
\end{flushright}
\vspace{20mm}

\begin{center}
{\large\bf
Non-ladder Extended Renormalization Group Analysis
\vskip2mm
of the Dynamical Chiral Symmetry Breaking
}
\vskip20mm
{
\large
Ken-Ichi Aoki,
Kaoru Takagi, 
Haruhiko Terao
\vskip2mm
and Masashi Tomoyose
}

\vskip10mm
{\large
Institute\ for\ Theoretical\ Physics, Kanazawa\ University,
\vskip2mm
Kanazawa\ 920-1192, Japan}

\end{center}

\vskip0mm
\centerline{\bf Abstract
}
\vskip7mm
\centerline{
\parbox{0.9\textwidth}{ 
The order parameters of dynamical chiral symmetry breaking in 
QCD, the dynamical mass of quarks and the chiral 
condensates, are evaluated by numerically solving the Non-Perturbative 
Renormalization Group (NPRG) equations. 
We employ an approximation scheme beyond ``the ladder'', that is,
beyond the (improved) ladder Schwinger-Dyson equations. 
The chiral condensates are enhanced compared with the ladder ones,
which is phenomenologically favorable. 
The gauge dependence of the order parameters is fairly reduced 
in this scheme. 
}}

\section{Introduction}

The dynamical chiral symmetry breaking plays an important role in the field of
particle physics.  
In particular, the hadron dynamics gives the most important example of the
dynamical chiral symmetry breaking phenomena.  
At present the low energy phenomenology of pions and kaons are well
understood by regarding them 
as the Nambu-Goldstone bosons of the dynamical chiral symmetry breaking
$SU(3)_{\rm\scriptsize L}\times SU(3)_{\rm\scriptsize R}
\rightarrow SU(3)_{\rm\scriptsize V}$ caused by 
the non-vanishing quark pair
condensate $\langle\bar qq\rangle\neq 0$ in QCD.  

There have been many studies about the dynamical chiral symmetry
breaking.  
The ladder Schwinger-Dyson (SD) equations in 
the Landau gauge have been mostly and extensively
used.\cite{NJL}-\cite{SDbeyond} \
In the strongly coupled QED, the chiral critical behaviors were explored
and the anomalous dimension of operator $\bar\psi\psi$ as well as 
the critical exponents of the phase transition, were
obtained.\cite{SDpioneer}-\cite{BHL01} \
The SD approach has been applied also to various models beyond the
standard model. \cite{SDmodel,topcondensation,BHL01} \
The chiral order parameters for QCD: the
dynamical mass of quarks, the chiral condensate and the decay constant 
of $\pi$ mesons $f_{\pi}$ were also calculated by using the improved
ladder approximation in the SD equations.\cite{SDimproved,SDQCD} \
They incorporate the running effects of the asymptotically free gauge
coupling constant in the ladder self consistent equations, 
thus called the improved ladder approximation, and
offered good results, even in quantities.  
However it should be noted that the improved ladder approximation has no 
theoretical justification, and it has been just an artificial ``model''.  
Therefore there has been no way to improve further
the ``improved ladder'' until
the Non-Perturbative Renormalization Group (NPRG)
\cite{WilKog,WegHou,Wilsonian01,Wilsonian02} analyses 
shed new light on it.\cite{KNZWsumi,KNZWWH01} \
Furthermore it has been shown that the ladder SD equations are plagued
with the strong gauge parameter dependence. \cite{SDdifficulty} \
Also it is difficult to proceed beyond the ladder approximation
so as to overcome this unpleasant problem.\cite{SDbeyond} \

The NPRG is also a powerful analytical technique for the
study of
non-perturbative phenomena in various fields of physics.\cite{NPRG01} \
The NPRG  has many favorable properties: 
The NPRG equations can be formulated
exactly and there are systematic methods of approximation.  
The exact NPRGs are given in the form of the nonlinear
functional differential equations for the Wilsonian effective action.  
Therefore, it is necessary to approximate them for the practical
calculations.  
We expand the Wilsonian effective action in powers of derivatives and 
truncate the series at a certain order.  
We often use the lowest order of this approximation, the so-called Local
Potential Approximation (LPA).\cite{LPA} \
Although the LPA looks like a very crude approximation, 
it gives very good results indeed,
for example, for the second order phase 
transition in $O(N)$ scalar field theories,
compared with other non-perturbative methods, 1/N-expansion and
$\epsilon$-expansion.\cite{KNZWscalar} \

In previous papers,\cite{KNZWsumi,KNZWWH01,KNZWWH02} we studied the
chiral critical behaviors mainly in the Abelian gauge theory with
strongly coupled massless fermions.  
We proposed a set of the gauge independent NPRG equations describing the 
chiral phase transition.  
Gauge independence is achieved by taking in the non-ladder type 
diagrams to the $\beta$ function of the four-fermi operators, 
as well as the ladder type ones.  
The gauge independent values of the critical exponent and of the
anomalous dimension of the mass operator $\bar\psi\psi$ were obtained.  
When we restrict the $\beta$ functions to the ``ladder parts'', 
the NPRG equations exactly reproduce the ladder SD results.  
It was analytically shown in the ladder approximation that 
not only the critical behaviors but also the order parameters 
are identical to the results of the SD equations. 
It is remarkable that this exact equivalence between the ladder part NPRG
and the ladder SD holds also for the improved ladder SD with the running 
gauge coupling constant.  
Thus our NPRG method has given for the first time a definite physical
meaning to the improved ladder SD, and now we know how to improve the
improved ladder, which we will challenge in this article.  

In Ref.~\cite{KNZWWH02} we proposed a new scheme of the NPRG equations 
incorporating the composite operators so as to definitely evaluate the order
parameters of the dynamical chiral symmetry breaking.  
This formulation was also applied to the non-Abelian gauge theories by
taking account of the asymptotically free running of the gauge coupling
constant.  
However the practical calculations of the order parameters were
demonstrated only in the ladder approximation.  
Therefore the results obtained there should depend on the gauge
parameter just as the ladder SD equations suffer.
Note here that our calculational method of incorporating the composite
operators should be distinguished from other schemes of constructing
the effective meson theory at some scale in QCD,
although they should be compared with each other. 
Also non-perturbative renormalization group analyses of QCD with
effective meson
components are done in a scope of the hadronic matter.\cite{MesonTheory}\  

In this paper we evaluate the chiral order parameters of QCD
in a new approximation scheme which is a minimal extension 
including the non-ladder type diagrams so as to compensate for the
serious gauge dependence of the ladder parts. 
The results obtained in this scheme, however, are not completely 
free of the gauge parameter dependence.  
There are still a few sources in our approximation causing the gauge
dependence.  
We examine the amount of the gauge dependence appeared in the order
parameters.  
It is found that the gauge dependence in the observable quantity, the
quark condensate $\langle\bar\psi\psi\rangle$, is fairly reduced
compared with the results in the ladder approximation.

The outline of this paper is as follows.  Section \ref{sec-NPRGandLPA}
is a brief review of the formalism upon which our work is based, the
method of NPRG and its approximation.  
We present our model in section
\ref{sec-NPRG equations for dynamical chiral symmetry breaking},
where we show how to treat the infrared divergences
occurring in the dynamical chiral symmetry breaking. 
In section \ref{sec-Compensation of the gauge dependence}  
we consider the origin of the gauge dependence in the ladder
approximation, and construct a new beyond the ladder approximation
which should reduce the gauge dependence. 
In section \ref{sec-Numerical Calculation and Results} we describe
the practical calculation procedures and we present numerical results. 
We discuss the remaining gauge dependence of the order parameters in our 
approximation in section \ref{sec-Issues of the gauge dependence}.  
Summary will be given in section \ref{sec-Summary and Discussion}.

\section{NPRG Equation and its Approximation}
\label{sec-NPRGandLPA}
There have been known several formulations
of the NPRG.\cite{WilKog,WegHou,Wilsonian01,Wilsonian02} \
In this paper we take the Wegner-Houghton (W-H)
equation.\cite{WegHou} \ First let us briefly review its formulation.
The starting point is the Euclidean path integral with the controlled 
momentum cutoff $\Lambda(t)=e^{-t}\Lambda_0$:
\begin{eqnarray}
 Z=\int^{\Lambda(t)} {\cal D}\phi \exp[-\Seff(\phi; t)],
\end{eqnarray}
where $\Seff$ is called the Wilsonian effective action.  
The NPRG equation describes how the Wilsonian effective action $\Seff$
should change as 
the higher momentum degrees of freedom are integrated out. 
It is obtained by reducing $\Lambda(t)$ infinitesimally with fixing the
partition function $Z$.  
Simultaneously we rescale the momentum variables
and the fields by cutoff
$\Lambda(t)$, since the change of the dimensionless quantities are of
our physical interest. 
We obtain the following differential equation,
\begin{eqnarray}
\frac{d\Seff[\phi; t]}{dt}
&=&D\Seff-
\int _{ |p|\leq 1}
 \frac{d^Dp}{(2\pi)^D} ~\phi_i(p)\left(
D_{\phi} -p^{\mu}\frac{\partial'}{\partial p^{\mu}}
\right)
\frac{\delta}{\delta \phi_i(p)}\Seff 
\nonumber\\
&-&\!\frac{1}{2dt}\int'\frac{d^Dp}{(2\pi)^D}
\left\{
\frac{\delta \Seff}{\delta\phi_i(p)}
\left(
 \frac{\delta^2 \Seff}{\delta\phi_i(p)\delta\phi_j(-p)}
\right)^{-1}\!\!\!
\frac{\delta \Seff}{\delta\phi_j(-p)}\right.
\nonumber\\
&&
\left.
\qquad\qquad\qquad\quad -\str\ln
\left(
 \frac{\delta^2 \Seff}{\delta\phi_i(p)\delta\phi_j(-p)}
\right)
\right\},
\end{eqnarray}
where $D$ is the space time dimension, $D_{\phi}$ is the dimension of
$\phi$ including its anomalous dimension, 
the second primed 
integral denotes integration over the infinitesimal shell modes of
momenta $e^{-dt} \leq p \leq 1$, 
and the prime in the
derivative indicates that it does not act on the $\delta$ function in
$\frac{\delta}{\delta \phi_p}\Seff $.  The subscript $i$ represents
every Lorentz and internal
symmetry indices.  
This equation is known as a
sharp cutoff version of the NPRG, and is called the
Wegner-Houghton (W-H) equation.\cite{WegHou} \
It is inevitable to approximate them for the practical calculations.  
We expand the Wilsonian effective action in powers of derivatives.  
We employ the Local Potential Approximation which is regarded as 
the lowest order of this derivative expansion. 
Any derivative couplings are dropped except for the
fixed kinetic terms,
\begin{eqnarray}
  \Seff&=&\int \frac{d^Dp}{(2\pi)^D}
  \left\{ \frac{1}{2}\phi_iK_{ij}(p)\phi_j
       +\Veff(\phi)\right\},
\end{eqnarray}
where $K_{ij}(p)$ is a matrix of the canonical kinetic terms,
and $\Veff$ is called the Wilsonian effective potential. 
As a simple example, we consider a theory of one scalar
$\varphi$ and one Dirac fermion $\psi$ and its conjugate $\bar\psi$.  
The matrix $K(p)$ in $(\phi, \psi, \bar\psi)$-space is written as
\begin{eqnarray}
K(p)=
 \left(
  \begin{array}{ccc}
   p^2 & 0       & 0 \\
   0   & 0       & -i\psla^T\\
   0   & -i\psla & 0
  \end{array}
 \right).
\end{eqnarray}
In this approximation the W-H equation is reduced to a nonlinear partial
differential equation for the Wilsonian effective potential
$\Veff(\phi, t)$,\cite{LPA}
\begin{eqnarray}
\frac{\partial\Veff(\phi; t)}{\partial t}
&=&D\Veff-
D'_{\phi} \phi_i
\frac{\partial \Veff}{\partial \phi_i}
+\frac{1}{2}\int'\! \frac{d^Dp}{(2\pi)^D}
\str\ln
\left(K_{ij}+
\frac{\partial^2 \Veff}{\partial\phi_i\partial\phi_j}
\right),
\label{e-LPAWH}
\end{eqnarray}
where $D'_{\phi}$ denotes the canonical dimension of field $\phi$.  
The anomalous dimension of field $\phi$ vanishes because the kinetic term
is not renormalized in the LPA.  
We should notice that any newly generated operators with derivatives
are ignored in this approximation. 
We take account of the generated operators which do not depend on the
external momenta.  
Definitely speaking, we evaluate amplitudes of such local non-derivative
operators by setting all external momenta to vanish. 

This partial differential equation may be solved numerically. 
However, actually it is not easy to get its solution
with enough precision. 
Besides it would not be practical for more complicated models. 
Therefore here we expand the
effective potential into polynomials in field $\phi$.  
By this approximation we solve a system of coupled ordinary differential
equations for the various coupling constants.  

Let us characterize these approximations from the viewpoint of the NPRG
formulation. 
The basic logic of approximation in the NPRG formalism is to restrict the
theory space to a subspace of the original full theory space.  
Namely the approximation in the NPRG formalism is to analyze the NPRG
equation projected onto a subspace, which is actually finite dimensional
so as to get results numerically within finite computation time. 
In order to improve the approximation, 
we enlarge the subspace, step by step, expecting the results will
converge to certain values. 
The above two approximations, the local potential approximation and
further the polynomial expansion, are just two consecutive steps of the
subspace projection. 
It should be noted that the ladder approximation itself can not be
regarded as projection to any subspace, and therefore it has some 
pathological features indeed.

Finally we should mention an intrinsic problem of the NPRG formulation. 
The momentum space cutoff is indispensable for almost any formulation of
the NPRG. 
It comes out that the NPRG does not manifestly respect the gauge
invariance. 
There have been several approaches for restoration of the gauge
invariance.\cite{Becchi01,MSTI,Morris02} \
Our purpose here is not to overcome this gauge invariance problem but to
improve the improved ladder approximation. 
Therefore we take a simple approximation scheme for the gauge interactions. 
We evaluate the $\beta$ function of the gauge coupling constant by 
the one-loop perturbative $\beta$ function. 
Of course the running of the gauge coupling constant is automatically
derived by the original NPRG equation. When we adopt a sub-theory space
with lower dimenaional operators, then the NPRG equation effectively
reproduces the perturbative renormalization group
equation.\cite{Wilsonian01,KNZWsumi} \
Therefore as the fist stage, we suppose this level of the small 
sub-theory space, and we adopt the running gauge coupling constant 
controlled by the perturbative $\beta$ function.  
We should notice that any newly generated operators including the gauge
fields are irrelevant in this scheme. 
Therefore the Faddeev-Popov ghosts are also irrelevant.

\section{NPRG equations for the dynamical chiral symmetry breaking}
\label{sec-NPRG equations for dynamical chiral symmetry breaking} 
Now we apply Eq. (\ref{e-LPAWH}) to QCD with three massless quarks.  
We take the local potential effective action,
\begin{eqnarray}
\Seff[\psi, \bar\psi, A_{\mu}; \ t] 
&=&
\int d^4x \left[\Veff(\psi,\bar\psi;\ t)
+\bar\psi (\dsla -g\Asla)\psi \right.
\nonumber\\ 
&&
\left.
+\frac{1}{4}(F^a_{\mu \nu})^2 
+ \frac{1}{2\alpha}(\partial_{\mu}A_{\mu}^a)^2\right],
\end{eqnarray}
where $\alpha$ is the gauge parameter, and $\psi$ denotes massless
triplet quarks.  
Here, as mentioned at the end of the previous section, 
the gauge coupling constant $g$ is supposed to follow the one-loop RG
equation. 
We start with the general form of the effective potential consistent 
with the chiral symmetry
$SU(3)_{\rm\scriptsize L}\times SU(3)_{\rm\scriptsize R}$ and the
parity. 
Let us first consider four-fermi operators. 
We regard the operators corresponding to the Fierz transformation as the
identical operators. 
Furthermore we shall not consider the flavor and/or color changing 
multi-fermi operators. 
Then there are two independent four-fermi operators in our theory space.  
\begin{eqnarray}
{\cal O}_1 &=& (\bar\psi \psi)^2+(\bar\psi i\gamma _5 \psi )^2
= -\frac{1}{2}\left\{
(\bar{\psi}\gamma_{\mu}\psi)^2-(\bar{\psi}\gamma_{5}\gamma_{\mu}\psi)^2
\right\}, \nonumber \\
{\cal O}_2&=&(\bar\psi\gamma_{\mu}\psi)^2+(\bar\psi \gamma_5
\gamma_{\mu} \psi )^2.
\end{eqnarray}
To approximate the NPRG equation, we specify a subspace in the full
theory space. 
Here we take a subspace spanned by polynomials in scalar
operator ${\cal O}_1$ only up to some maximum power $nmax$,
\begin{eqnarray}
\Veff(\psi, \bar\psi; \ t) 
&=& \sum_n^{nmax}\frac{G_{2n}( t )}{n}
\left[(\bar\psi\psi)^2 +(\bar\psi i\gamma_5\psi)^2\right]^n.
\label{e-naiveexpansion}
\end{eqnarray}

\begin{figure}[htb]
 \begin{minipage}{65mm}
\vskip1mm
  \epsfxsize=65mm
  \centerline{\epsfbox{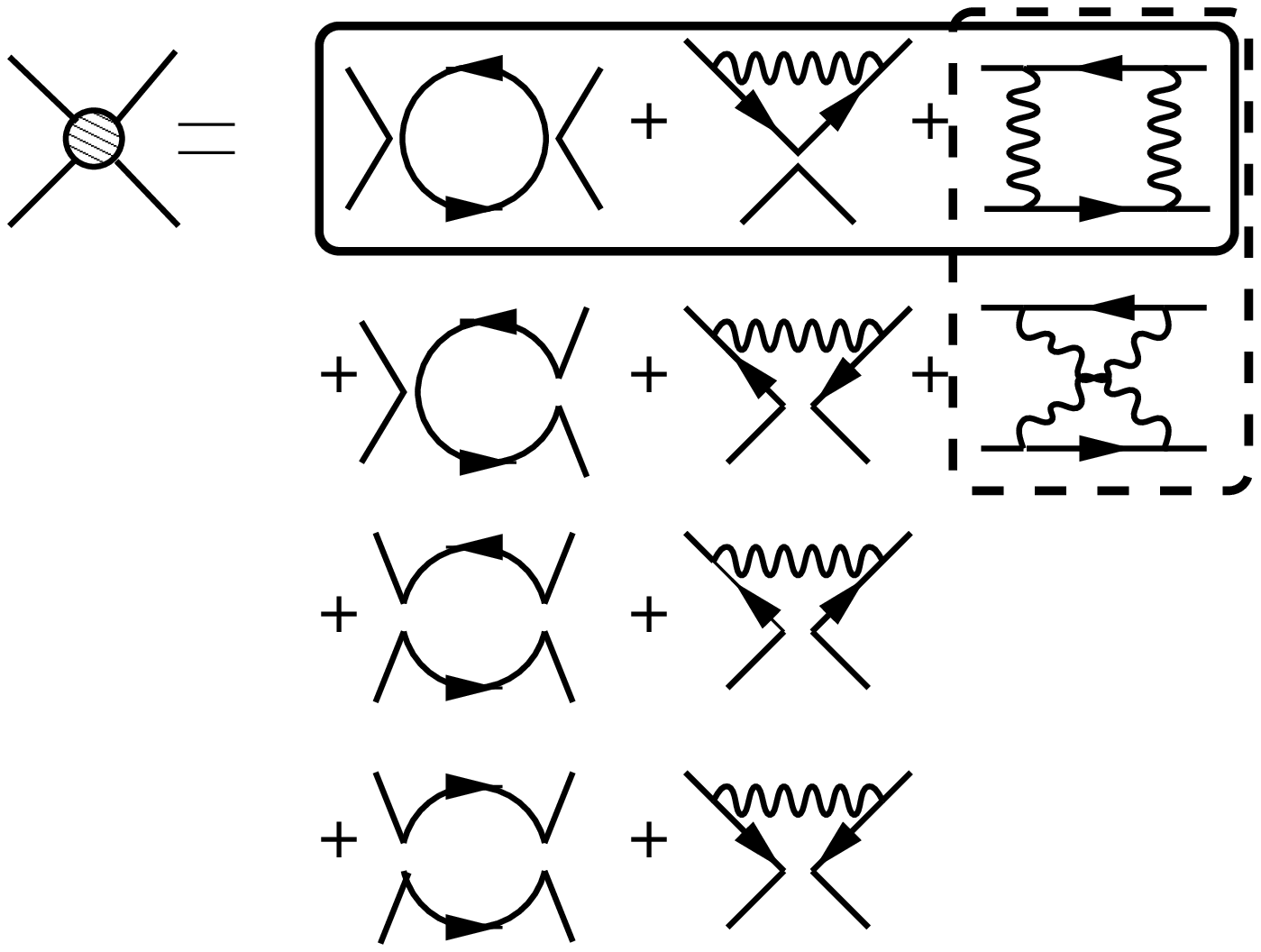}}
   \caption{\footnotesize 
    The NPRG $\beta$ function of the four-fermi operators. 
    The diagrams in the solid box correspond to the ladder part, 
    and the diagrams in the dashed box are crucial for the gauge 
    independence.  
    Possible other diagrams are ignored in our approximation. 
    \label{4fermidiagram}}
 \end{minipage}
\hfil
 \begin{minipage}{65mm}
   \epsfxsize=65mm
  \centerline{\epsfbox{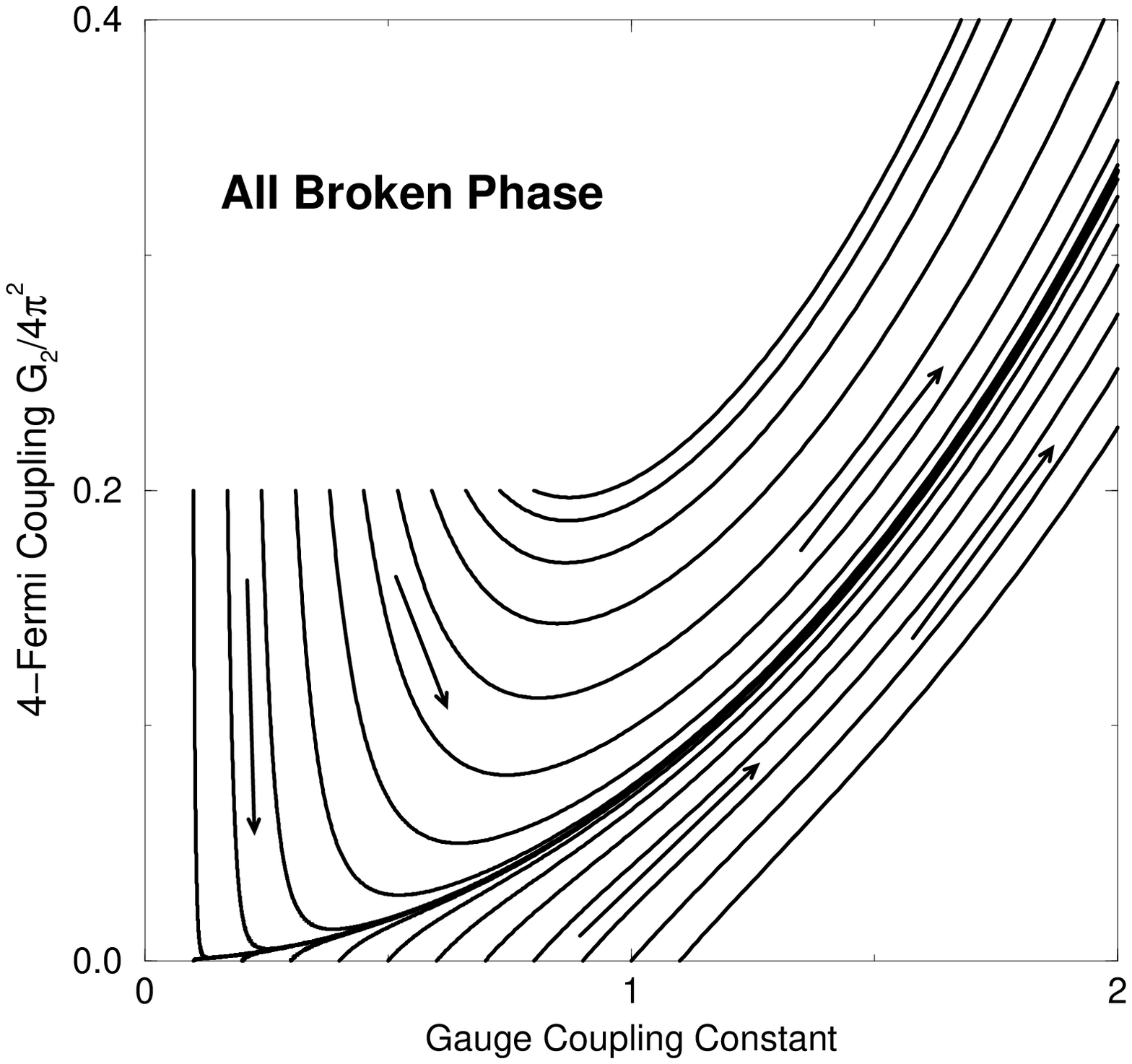}} 
   \caption{\footnotesize 
   The flow diagram in the $G_2-g$ plane in QCD.  
   The phase structure does not appear, and all flows diverge at a 
   certain finite scale. 
   \label{QCDflow}}
 \end{minipage}
\end{figure}

The NPRG equation for the scalar four-fermi operator is obtained from
the diagrams in Fig. \ref{4fermidiagram},
\begin{equation}
 \frac{dG_2}{dt}=
   -2G_2
   +\frac{1}{2\pi^2}(G_2)^2
   +\frac{3 g^2}{4\pi^2}G_2
   +\frac{9 g^4}{32\pi^2}~.
\label{e-4-fermi-naive-ladder}
\end{equation}
Generally the NPRG equation is a set of coupled equations of various
operator. 
In this case, however, the NPRG $\beta$ function of four-fermi operator
${\cal O}_1$ consists of ${\cal O}_1$ itself and the gauge coupling
constant $g$. 
Therefore the NPRG equation for ${\cal O}_1$ operators can be solved
without recourse to other higher multi-fermi operators. 
When ignoring the running of the gauge coupling constant,
there is a fixed point for the flow of ${\cal O}_1$ operator given by
the zero of the right-handed side of Eq. (\ref{e-4-fermi-naive-ladder}),
which is nothing but the critical point of the dynamical chiral symmetry
breaking,
and we have two-phase structure of the standard ferromagnet type phase
transition.\cite{KNZWsumi,KNZWWH01} \
The strong coupling phase is the symmetry breaking phase, where the
four-fermi coupling constant diverges at a finite scale.

We make the gauge coupling constant run according to the asymptotically
free $\beta$ function. 
The flow diagram in the $G_2-g$ plane in QCD is depicted in
Fig. \ref{QCDflow}.  
There appears no phase boundary.  All the flows diverge at a certain
finite scale. 
Also we see the renormalized trajectory in the flow diagram which assures 
that the bare four-fermi interactions are irrelevant to the infrared
physics. 
This behavior suggests that the entire region is supposed to be in the
broken phase of the chiral symmetry. 
This is due to the infrared slavery behavior of the QCD gauge coupling
constant and it is believed to hold naturally. 

Renormalization group flows have been also analyzed ~by the SD equation 
method\cite{KSY}, where fixing the quark mass obtained, the relation 
between the four-fermi coupling constant and the gauge coupling constant
is calculated assuming the cutoff dependence of the gauge coupling
constant. 
This procedure is justified within the SD formalism and it actually
gave something resembling to the results in Fig. \ref{QCDflow}.
However there are critical differences between these two calculations.
The main difference comes from the fact that in the NPRG formalism 
the bare four-fermi interactions are turned out to be irrelevant,
while in the SD formalism there is no mechanism to automatically 
generate effective four-fermi interactions by the gluon exchanges.

Correspondence between the divergence of the four-fermi operator and the
dynamical chiral symmetry breaking is not trivial. 
All the coupling constants in the polynomial expansion
keep growing in the infrared and diverge at some finite scale. 
This corresponds to the fact that
at this scale the Wilsonian effective potential exhibits a
non-analytical behavior at its origin,
which is actually observed as a jump of the first derivative by 
direct analysis of the Wilsonian effective potential using the partial
differential equation. 
This singularity at the origin clearly shows the spontaneous symmetry
breakdown since it assures the non-vanishing magnetization at zero
external field limit.

This singular behavior invalidates the renormalization group calculation 
of the evolution of the effective potential with polynomial expansion at 
the origin. 
Introducing a composite operator
$\phi$ corresponding to the order parameter $\langle\bar\psi\psi\rangle$
enables us to carry out the calculation of the Wilsonian effective
potential even in the infrared region.\cite{KNZWWH02} \
Our theory space is extended to include the composite operator
$\phi$. 
First we introduce the composite operator $\phi$ as an auxiliary field
in the original path integral without changing the dynamics. 
The partition function $Z$ in this extended theory space is written as
\begin{eqnarray} 
Z&=&
\int {\cal D}\psi{\cal D}\bar\psi {\cal D}A
\exp \left[\,-\int d^4x\ {\cal L}_{\rm org}\:[\psi ,\bar\psi ,A\;]
\; \right]
\nonumber\\
&=&
\int {\cal D}\psi{\cal D}\bar\psi {\cal D}A{{\cal D}\phi} 
\exp \left[\,-\int d^4x\left\{ 
{\cal L}_{\rm org}\:[\psi ,\bar\psi ,A\;]
+\frac{1}{2}(\phi-y\bar\psi \psi )^2 
\right\}\; \right].
\end{eqnarray}%
Here we abbreviated the counter part $(\bar\psi i\gamma_5\psi)^2$ 
for simplicity
and it should be considered that the original chiral symmetry is 
kept actually.
Then the bare lagrangian is modified as follows: 
\begin{eqnarray} 
{\cal L}&=& 
{\cal L}_{\rm org}\:[\psi ,\bar\psi ,A; \ t]
+\frac{1}{2}(\phi-y\bar\psi \psi )^2 
\nonumber\\
&=&
 \frac{1}{4}F^a_{\mu \nu}F^a_{\mu \nu}
+\frac{1}{2\alpha}(\partial_{\mu}A_{\mu}^a)^2
+\bar\psi(\dsla -g\Asla -y\phi)\psi
+\frac{1}{2}\phi^2 +\frac{y^2}{2}(\bar\psi\psi)^2.  
\label{e-InitialPotential}
\end{eqnarray}%
Even in case of omitting the chiral symmetric representation,
this lagrangian has the discrete chiral symmetry: 
\begin{eqnarray}
 \bar\psi \rightarrow \gamma_5 \psi, \quad
 \psi \rightarrow -\bar\psi\gamma_5, \quad
 \phi \rightarrow -\phi.
\end{eqnarray}
Then we may analyze the NPRG equations for the following Wilsonian
effective potential: 
\begin{eqnarray} 
\Leff
&=& 
\frac{1}{4}F^a_{\mu \nu}F^a_{\mu \nu}+\bar\psi
(\dsla -g\Asla)\psi+\Veff(\phi, \sigma; \ t),\\
 \Veff(\phi, \sigma; \ t)
&=&
 \tilde G_0(\phi; \ t)+
 \sum^{nmax}_n\frac{1}{n}\tilde G_n(\phi; \ t)\ \sigma^n
\nonumber\\
&=&
\tilde G_0(\phi; \ t)+\tilde G_1(\phi; \ t)\sigma+
\frac{1}{2}\tilde G_2(\phi; \ t)\sigma^2+\cdots,
\label{e-EffectivePotential-ExtendedExpantion}
\end{eqnarray}
where the notation $\sigma = \bar\psi\psi$ is introduced. 
Note again that we are actually working with the chiral symmetric
effective potential and we take a particular direction of the scalar
operator $\sigma$ condensation to get the effective potential
represented by Eq. (\ref{e-EffectivePotential-ExtendedExpantion}). 
Since we do not consider the propagation of composite operator modes,
Eq.(\ref{e-EffectivePotential-ExtendedExpantion}) is enough to define the
NPRG evolution of the chiral symmetric system.

In this formalism, it was shown that the chiral condensate
$\langle\bar\psi\psi\rangle$ is proportional to the minimum position
of the scalar 
potential $\tilde G_0(\phi)$, denoted by $\langle\phi\rangle$,
which is the order parameter of the dynamical chiral symmetry breaking.  
\begin{eqnarray}
 \langle\bar\psi\psi\rangle&=&\frac{1}{y}\langle\phi\rangle,
 \label{e-condensate}
\end{eqnarray}
Then the dynamical mass of quarks $\Sigma(0)$ is given by
\begin{eqnarray}
 \Sigma(0)&=&\left.\frac{\partial\Veff}{\partial \sigma}
             \right|\hspace*{-2mm}_{\tiny
                      \begin{array}{l}
                       \phi=\langle\phi\rangle\\
                       \sigma=0
                      \end{array}}
           =\tilde G_1(\phi=\langle\phi\rangle)
 \label{e-Sigma(0)}
\end{eqnarray}
In the usual argument of introducing the auxiliary field,
the four-fermi interaction is removed away from the action by tuning the 
Yukawa coupling constant $y$.  
Then this action may be regarded as the gauged Yukawa system with
the compositeness condition.\cite{BHL01} \
There are several studies of the dynamical chiral symmetry breaking by
using this realization.  
However, we should notice that our purpose of introducing the
auxiliary field $\phi$ is absolutely
not to eliminate the four-fermi interaction.
Also our results obtained by Eqs.(\ref{e-condensate}) and 
(\ref{e-Sigma(0)}) are turned out to be 
independent of the
Yukawa coupling constant $y$,\cite{KNZWWH02} \
since it should not change the dynamics at all.

\section{Compensation of the gauge dependence}
\label{sec-Compensation of the gauge dependence}
As noted before, the ladder part NPRG exactly reproduces the
results obtained by the ladder SD equations.  
Namely the results by the ladder part NPRG depend on the gauge
parameter $\alpha$ strongly, like the ladder SD approaches.  
In order to improve the gauge dependence, we must proceed beyond the
ladder approximation.  
So it is required to develop a non-ladder extended approximation 
in the course of the systematic approximation of NPRG.  

\begin{figure}[htb]
\centerline{\epsfxsize=0.9\textwidth \epsfbox{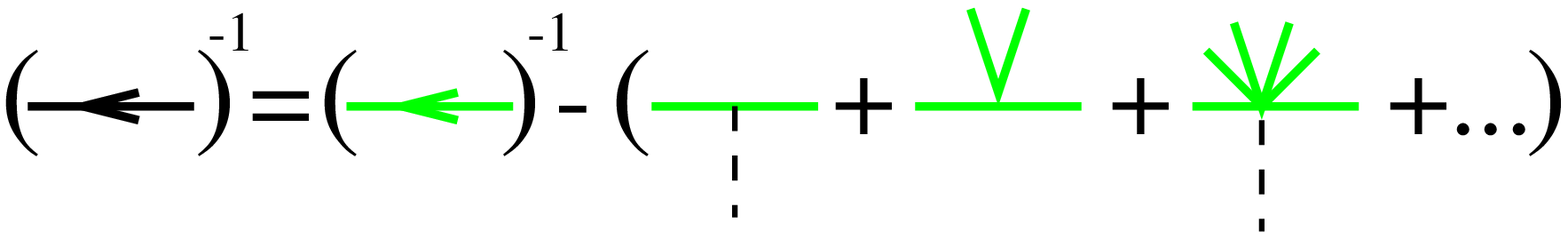}}
\centerline{ \parbox{0.9\textwidth}{ 
 \caption{\footnotesize 
  The ``massive'' quark propagator.  
  The deep full line in the left-handed side is a ``massive'' fermion 
  propagator.  The pale full lines are massless fermion operators and
  the dashed lines are the auxiliary fields $\phi$.
 \label{massivepropa}}}}
\end{figure}

\begin{figure}[htb]
 \begin{minipage}{20mm}
\vspace*{-9mm}
  \begin{eqnarray*}
   {\huge \frac{d \Veff}{dt}= }
  \end{eqnarray*}
 \end{minipage}
\hspace*{-2mm}
 \begin{minipage}{22mm}
  \hfil
  \epsfxsize=20mm
  \epsfbox{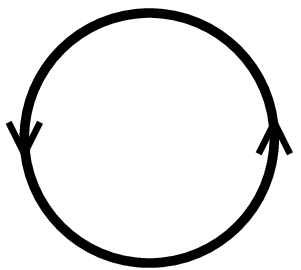}
 \end{minipage}
 \begin{minipage}{9mm}
  \begin{center}
   {\huge $+$}
  \end{center}
 \end{minipage}
 \begin{minipage}{32mm}
  \epsfxsize=32mm
  \epsfbox{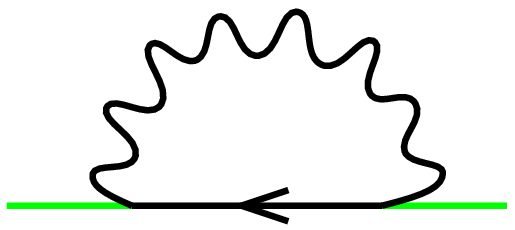}
 \end{minipage}
\hspace*{1mm}
 \begin{minipage}{9mm}
   {\huge $+$ }
 \end{minipage}
 \begin{minipage}{31mm}
  \epsfxsize=31mm
  \epsfbox{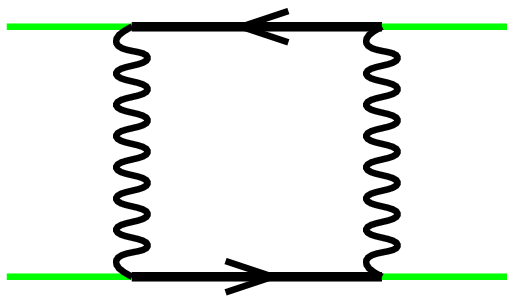}
 \end{minipage}
 \vskip4mm
\hspace*{2mm}
 \begin{minipage}{9mm}
   {\huge $+$ }
 \end{minipage}
 \begin{minipage}{29mm}
  \epsfxsize=28mm
  \epsfbox{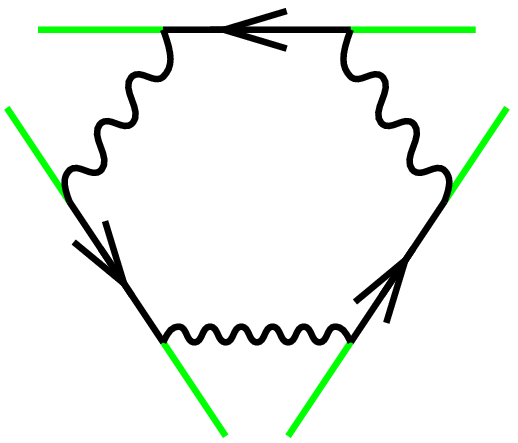}
 \end{minipage}
\hspace*{1mm}
 \begin{minipage}{12mm}
  {\huge $+$ }
 \end{minipage}
 \begin{minipage}{25mm}
  \epsfxsize=23mm
  \epsfbox{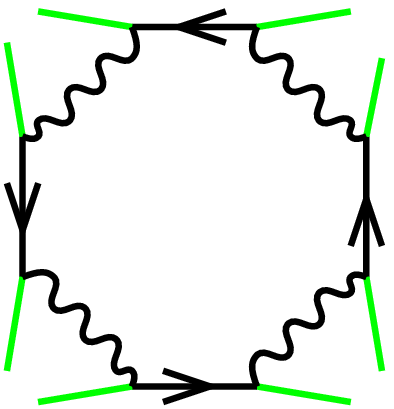}
 \end{minipage}
\hspace*{3mm}
 \begin{minipage}{20mm}
  \hfil
   {\huge$+ \cdots$}
 \end{minipage}\\ \\  
\centerline{ \parbox{0.9\textwidth}{ 
 \caption{\footnotesize 
  The ladder part $\beta$ function. 
  These diagrams do not contain any crossed ladder type 
  diagrams. 
  The wavy lines are gluons, and the deep full lines are the ``massive'' 
  quarks, and the pale full lines are external quark operators. 
 \label{laddersum}}}}
\vskip7mm%
\end{figure}

To begin with, we discuss the origin of the gauge dependence in
the ladder approximation.  
Let us consider a set of diagrams summed up by the ladder approximation.  
For that purpose we define the ``massive'' quark propagator,
\begin{eqnarray}
 \frac{1}{i\psla-m(\phi,\sigma)}, 
\end{eqnarray}
where
\begin{eqnarray}
m(\phi, \sigma)
 &\equiv&
\frac{\partial V}{\partial\sigma}
=\frac{\partial}{\partial\sigma}
 \sum_{n=1}^{nmax}\frac{\tilde G_n(\phi)}{n}\sigma^n
=
 \sum_{n=1}^{nmax}\tilde G_n(\phi)\ \sigma^{n-1}
\nonumber\\
\nonumber\\
&=&
 \tilde G_1(\phi)+\tilde G_2(\phi)\sigma+\tilde G_3(\phi)\sigma^2
 +\tilde G_4(\phi)\sigma^3+\cdots.
\end{eqnarray}
Using the Feynman diagrams the ``massive'' quark propagator is
represented as shown in Fig. \ref{massivepropa}. 
The ladder part $\beta$ functions are defined
by summing up a set
of diagrams which do not contain any crossed ladder type diagrams
(Fig. \ref{laddersum}). 

We now consider the Abelian Ward-Takahashi (WT) identities
assuring the gauge independence. 
To satisfy WT identities we must sum over the
diagrams for the S-matrix at any given order.  
When the gauge boson is inserted at a certain point along the fermion line,
we must sum over all possible insertion points.  
Let us consider the simple case of WT identity involving two
fermions and two gauge bosons in $g^2$ order.  
 \begin{figure}[htb]
\centerline{
  \begin{minipage}{50mm}
   \epsfxsize=45mm
   \epsfbox{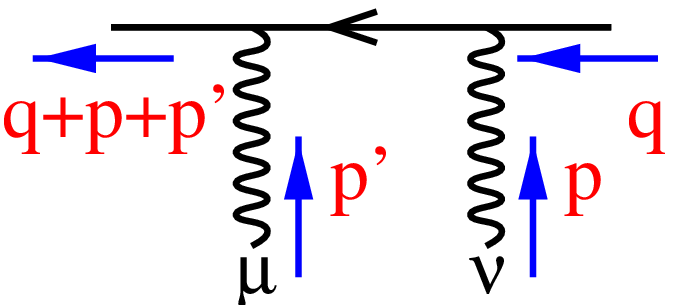}
  \end{minipage}
  \begin{minipage}{20mm}
   \vskip-6mm
     \centerline{\LARGE +}
  \end{minipage}
  \begin{minipage}{50mm}
   \epsfxsize=45mm
   \epsfbox{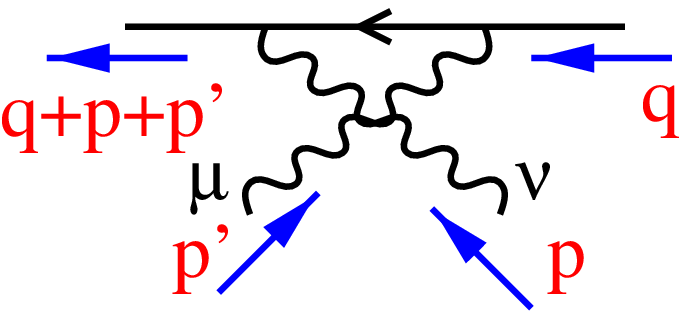}
  \end{minipage}}
\centerline{ \parbox{0.9\textwidth}{ 
\caption{\footnotesize 
 The set of diagrams satisfying the Ward-Takahashi identity which
 involves two fermions and two gauge bosons. 
 The wavy lines are gauge bosons and the full lines are fermions.  
 \label{QEDWTI}}}}
 \end{figure}
The sum of diagrams in Fig. \ref{QEDWTI} is gauge independent at on-shell.  
The right diagram of Fig. \ref{QEDWTI},
called the crossed diagram,
is not involved in the ladder approximation. 
This is one of the reason for the strong gauge dependence in the ladder
approximation. 
Actually as seen in Fig. \ref{4fermidiagram},
adding the crossed box diagram to the $\beta$ function has wiped out the 
gauge dependence of the critical behaviors.\cite{KNZWsumi}\

\begin{figure}[ht]
\centerline{
  \begin{minipage}{30mm}
   \epsfxsize=30mm
   \epsfbox{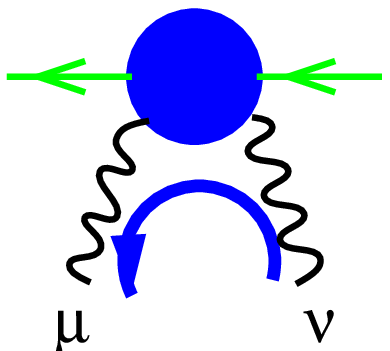}
  \end{minipage}
  \begin{minipage}{20mm}
   \begin{center}
    {\LARGE =}
   \end{center}
  \end{minipage}
  \begin{minipage}{30mm}
   \epsfxsize=30mm
   \epsfbox{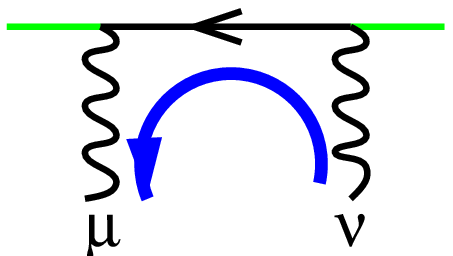}
  \end{minipage}
  \begin{minipage}{20mm}
   \begin{center}
    {\LARGE +}
   \end{center}
  \end{minipage}
  \begin{minipage}{30mm}
   \epsfxsize=30mm
   \epsfbox{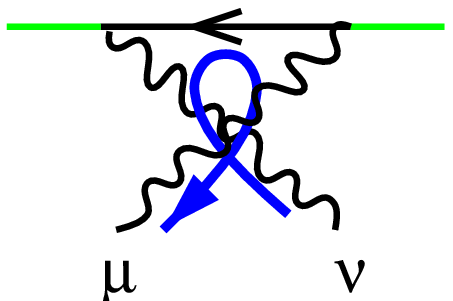}
  \end{minipage}}
\centerline{ \parbox{0.9\textwidth}{ 
  \caption{\footnotesize 
   The corrected vertex.
   The wavy lines are gluons, 
   and the deep full lines are ``massive'' quarks, 
   and the pale full lines are external quark operators. 
   The curved arrows denote the direction of the shell-mode momentum $p$.  
   \label{correctedvertex}}}}
 \end{figure}

\begin{figure}[htb]
\begin{eqnarray*}
\frac{dV}{dt}
=
\frac{d}{dt}
\left\{ \tilde G_0
       +\tilde G_1(\bar\psi\psi)
       +\frac{1}{2}\tilde G_2(\bar\psi\psi)^2
       +\frac{1}{3}\tilde G_3(\bar\psi\psi)^3
       +\frac{1}{4}\tilde G_4(\bar\psi\psi)^4  +\cdots
\right\}
\end{eqnarray*}%
\hspace*{2mm}
 \begin{minipage}{14mm}
  {\huge
   $=$
  }
 \end{minipage}
 \begin{minipage}{22mm}
  \epsfxsize=20mm
  \epsfbox{0gauge.eps}
 \end{minipage}
\hspace*{3mm}
 \begin{minipage}{9mm}
  \centerline{
   {\huge $+$}}
 \end{minipage}
 \begin{minipage}{40mm}
  \epsfxsize=37mm
  \epsfbox{1gauge6.eps}
 \end{minipage}
\hspace*{1mm}
 \begin{minipage}{15mm}
  {\huge $+~~\frac{1}{2}$}
 \end{minipage}
 \begin{minipage}{20mm}
  \epsfxsize=20mm
  \epsfbox{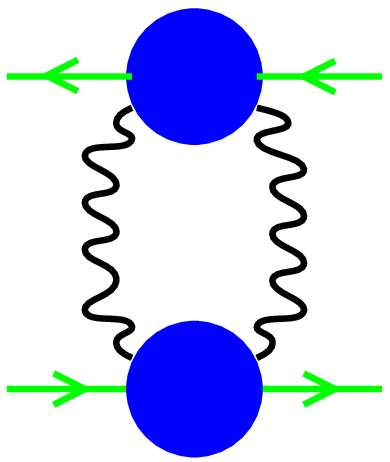}
 \end{minipage}
 \vskip1mm
\hspace*{2mm}
 \begin{minipage}{9mm}
  {\huge $+$ }
 \end{minipage}
 \begin{minipage}{32mm}
  \epsfxsize=30mm
  \epsfbox{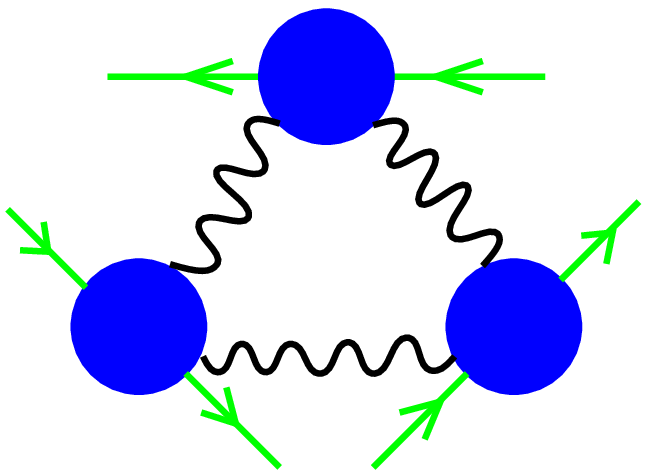}
 \end{minipage}
\hspace*{1mm}
 \begin{minipage}{12mm}
  {\huge $+$ }
 \end{minipage}
 \begin{minipage}{25mm}
  \epsfxsize=25mm
  \epsfbox{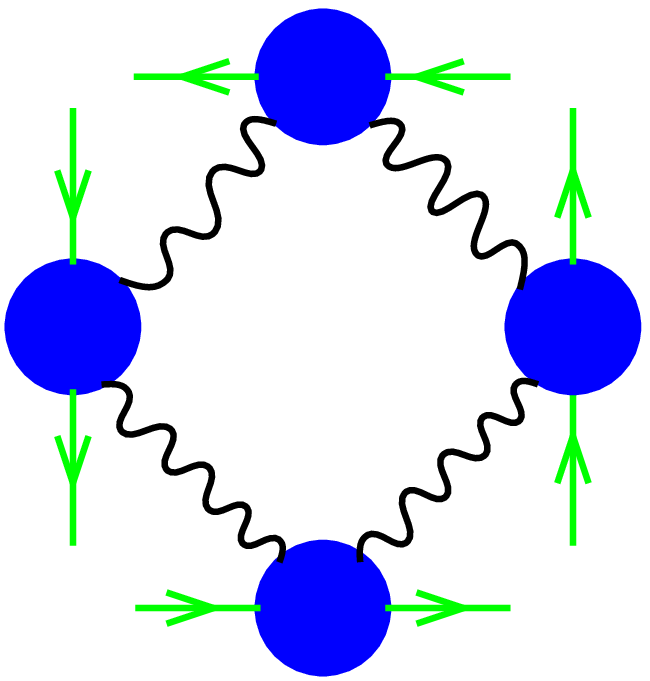}
 \end{minipage}
\hspace*{7mm}
 \begin{minipage}{20mm}
  {\huge$+ \cdots$}
 \end{minipage}\\
\centerline{\parbox{0.9\textwidth}{ 
 \caption{\footnotesize 
  The $\beta$ function in our new approximation.
  The wavy lines are gluons, and the deep full lines are ``massive'' 
  quarks,    and the pale full lines are external quark operators.
\label{nonladdersum}}}}
\end{figure}

Now we consider to generalize the crossed box diagrams in 
Fig. \ref{4fermidiagram} for the $\beta$ functions of the higher
multi-fermi operators. 
First we define a corrected vertex,
\begin{eqnarray}
 \frac{-2g^2}{p^2 + m^2 }
  \left[
    i\smp{\alpha}\epsilon^{\mu\nu\alpha\beta}\gammafive\gamma_{\beta}
   +m\Sg{\mu\nu}
  \right],
\label{e-correctedvertex}
\end{eqnarray}%
which is composed of ``two'' diagrams, ladder and crossed,
using the ``massive'' quark propagator (Fig. \ref{correctedvertex}).  
Therefore the corrected vertex itself
comprises an infinite number of diagrams.  
Then we replace double vertices in Fig. \ref{laddersum}
with the corrected vertex and sum up the
diagrams just as the ladder part (Fig. \ref{nonladdersum}).

Just as in Fig. \ref{laddersum}, every diagram in
Fig. \ref{nonladdersum} contributes to an infinite number of coupling
constants in the polynomial expansion. 
Accordingly, a certain diagram with $n$ external quarks contributes to
the $\beta$ functions of coupling constants
with more than $n$ external quarks.   
For instance, the $\beta$ function of mass $\tilde G_1$
(the generalized Yukawa coupling) is given by
the first and the second diagrams in Fig. \ref{nonladdersum}.  
Also the $\beta$ function of the four-fermi operator comes from
the first, the second and the third diagrams, etc.  
Note that the third diagram has a symmetry factor of 1/2.

\begin{figure}[htb]
\centerline{ 
\epsfxsize=25mm
\epsfbox{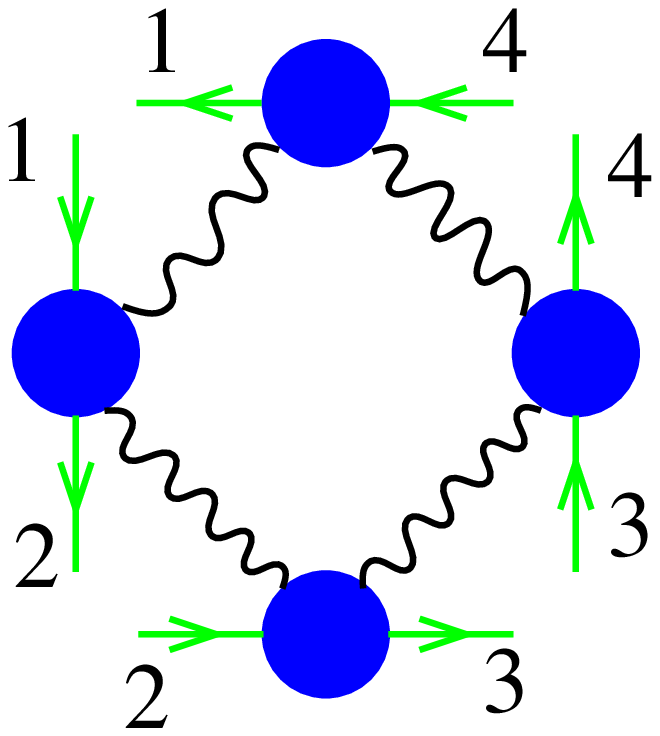}}
\centerline{ \parbox{0.9\textwidth}{ 
 \caption{\footnotesize 
  The newly generated $g^8$ operator.  
  The numbers denote the suffixes of the quark fields 
  (see Eq. (\ref{e-newlygeneratedoperator})).
 \label{newlygeneratedoperator}}}}
\end{figure}

\begin{figure}[h]
\begin{minipage}{65mm}
\hfill
\epsfxsize=24mm
\epsfbox{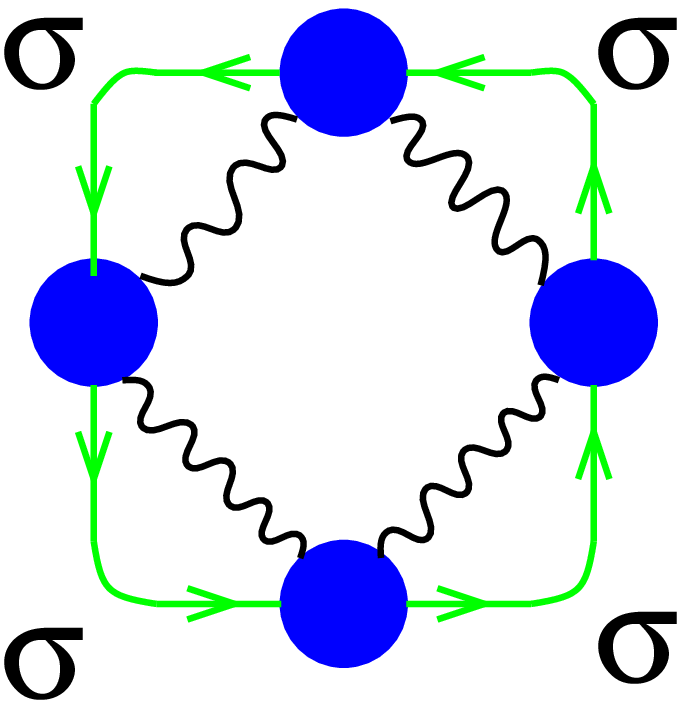}
\end{minipage}
\hfill
\begin{minipage}{65mm}
\epsfxsize=32mm
\epsfbox{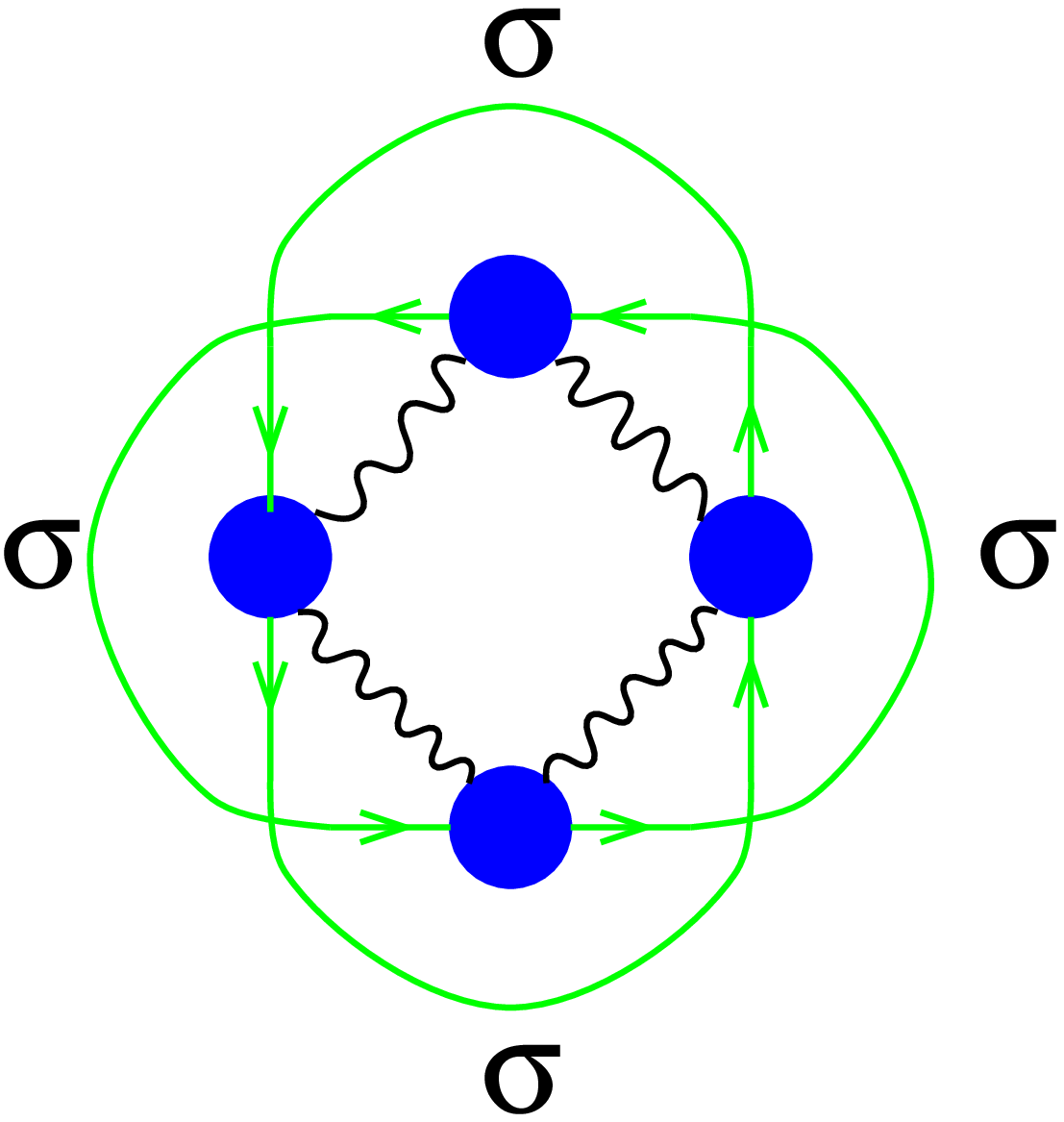}
\hfill
\end{minipage}
\centerline{ \parbox{0.9\textwidth}{ 
 \caption{\footnotesize 
  The way of picking up $\sigma$.  
  The left figure corresponds to the ladder part approximation, 
  when the corrected vertices are replaced with the ladder type. 
  We ignore the way of picking up $\sigma$ like the right figure. 
 \label{pickupsigma}}}}
\end{figure}

Now we discuss the way of ``projection'' of the newly generated
operators onto our target subspace. 
Here we consider the effective potential $\Veff$ composed of the
polynomials only in the scalar operator ${\cal O}_1$.  
Therefore we pick up only the parts proportional to $({\bf 1})^n$ in the
spinor space from the generated total operators.  
For example, let us consider $g^8$ terms in the eight-fermi beta
function (Fig. \ref{newlygeneratedoperator}).  
\noindent 
The form of the generated operators are in general,
\begin{eqnarray}
&&
\sum_{a,b,c,d}{\cal C}_{abcd}
 (\bar\psi_1\Gamma^{a}\psi_1)\times 
 (\bar\psi_2\Gamma^{b}\psi_2)\times
 (\bar\psi_3\Gamma^{c}\psi_3)\times
 (\bar\psi_4\Gamma^{d}\psi_4),
\label{e-newlygeneratedoperator}
\end{eqnarray}
where $\Gamma$'s are the 16 independent matrices for spinor indices
and ${\cal C}_{abcd}$ are amplitudes. 
We should consider all the parts proportional to 
$(\bar\psi{\bf 1}\psi)^4$ from the generated operators.  
Instead of picking up all of them, however,
we take only one simple combination of $\Gamma$'s,
\begin{eqnarray}
&&
 (\bar\psi_1{\bf 1}\psi_1)\times 
 (\bar\psi_2{\bf 1}\psi_2)\times
 (\bar\psi_3{\bf 1}\psi_3)\times
 (\bar\psi_4{\bf 1}\psi_4).  
\label{e-projectionto1}
\end{eqnarray}
This part is represented in the left figure in Fig. \ref{pickupsigma}. 
Of course there are another combinations of $\Gamma$'s contributing to 
$(\bar\psi{\bf 1}\psi)^4$ term,
for example, contraction shown in the right figure in
Fig. \ref{pickupsigma}, which is omitted. 
We adopt this approximation since
the left figure in Fig. \ref{pickupsigma} exactly coincides with the 
ladder part approximation, 
when the corrected vertices are replaced with the ladder type.
We now challenge the minimal extension of the ladder approximation. 
We should mention that this restriction of the
way of picking up $\sigma$ fields does not correspond to a general
procedure of projection in the systematic approximation method of NPRG,
and it might cause a problem. 

Now let us consider the non-Abelian effects. 
The quarks belong to the 3 dimensional representation.  
For example, $g^4$ term in the four-fermi $\beta$ function from the ladder
type diagram (Fig. \ref{box}) is identical to 
the Abelian estimate, except that we must add the color Casimir
eigenvalue,
\vskip2mm
\refstepcounter{figure}
\label{box}
\centerline{
 \begin{minipage}{40mm}
\epsfxsize=40mm
\epsfbox{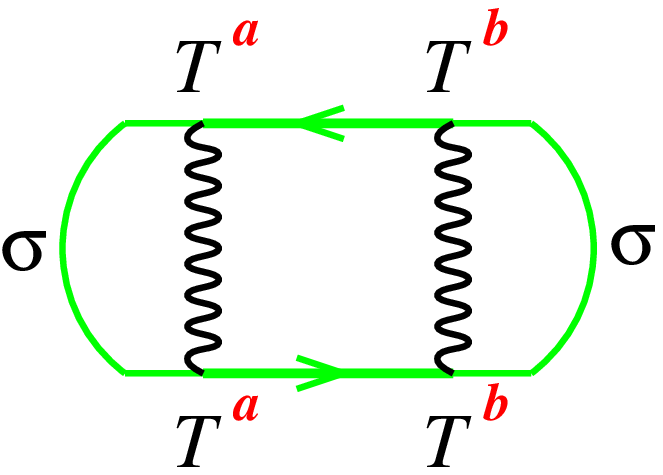}
 \end{minipage}
 \begin{minipage}{80mm}
\Large ${\Rightarrow}~ \Tr~ (T^aT^bT^bT^a)~=~3{C_2(3)^2}.$
 \end{minipage}}
\vspace*{2mm}
\centerline{\parbox{0.9\textwidth}{\footnotesize
Fig. \ref{box}. The gauge group factor of the box diagram 
in the four-fermi $\beta$ function.  
}}
\vskip5mm
\noindent
As for the crossed ladder type diagrams (Fig. \ref{crossed}),
there appears the commutator 
term $i f^{abc}$ in addition to the Casimir term. 
Here we simply ignore the commutator term $if^{abc}$ and take account of
the $C_2(3)^2$ term only. 
\vskip5mm
\refstepcounter{figure}
\label{crossed}
\centerline{
 \begin{minipage}{40mm}
\epsfxsize=40mm
\epsfbox{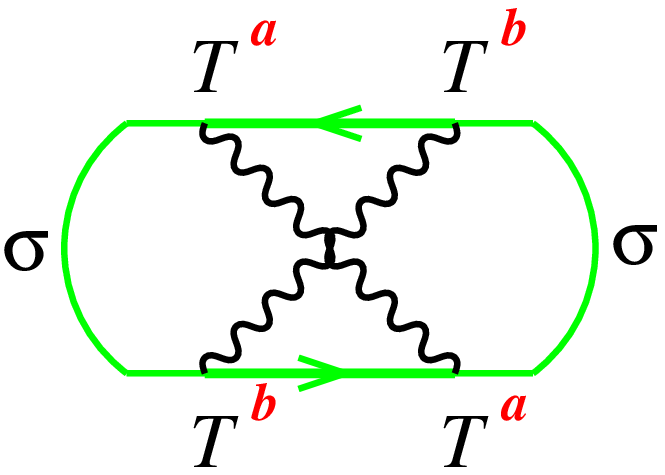}
 \end{minipage}
 \begin{minipage}{100mm}
\Large
${\Rightarrow} \Tr (T^aT^bT^aT^b) 
=3 C_2(3)^2 +if^{abc} \Tr(T^aT^bT^c).$
\end{minipage}}
\vskip5mm
\centerline{\parbox{0.9\textwidth}{\footnotesize
 Fig. \ref{crossed}. The gauge group factor of the crossed box diagram 
 in the four-fermi $\beta$ function.  }}
\vskip5mm
\noindent
Of course this is a violent truncation which might
break the gauge independence and it will be discussed later. 
Due to this additional approximation the non-Abelian nature is absorbed
into the Casimir factor which defines the effective gauge coupling
constant $C_2(3)g$.
With this effective gauge coupling constant, the $\beta$ function of the 
effective potential in QCD is evaluated just as in the Abelian case.

Now we are able to write down general formulae of  $g^{2n}$ terms 
in the NPRG equations as follows: 
\begin{figure}[ht]
 \begin{minipage}{42mm}
  \epsfxsize=40mm
  \epsfbox{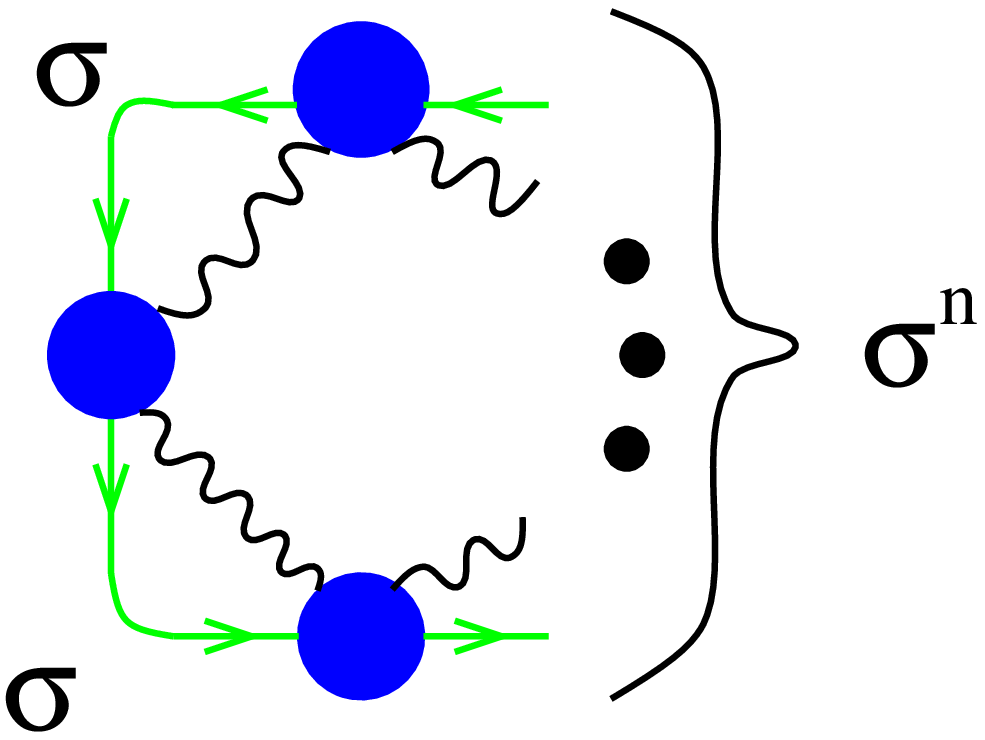}
 \end{minipage}
\end{figure}
\begin{eqnarray}
 &=&
 -\frac{1}{8\pi^2}\frac{g^{2n}}{2^{n-2}\ n!}\ 
 \frac{1}{(1+m^2)^n}\ 
 \frac{1}{1+\delta_{n,1}+\delta_{n,2}}
 \nonumber\\
 &&
 \times
 \left[
 m^n(3+{\alpha^n})+\sum_{k=1}^{[\frac{n}{2}]}(-)^k
 \frac{n!}{(2k)!\ (n-2k)!}
 (2+4^k) m^{n-2k}
 \right]
 \sigma^n.
\label{e-GeneralFoumula}
\end{eqnarray}
For example, the $\beta$ function for the four-fermi operator reads
\begin{eqnarray}
 \frac{d\tilde G_2}{dt}&=&
-2\tilde G_2
+\frac{1}{2\pi^2}\ \frac{\tilde G_2^2}{1+\tilde G_1^2}
 \left[
  1-\frac{2\tilde G_1^2}{1+\tilde G_1^2}
 \right]
\label{e-4fermiNL}
\\
&&
+\frac{g^2}{4\pi^2}\ 
 \frac{3+{\alpha}}{1+\tilde G_1^2}
 \left[
  \tilde G_2-\frac{2\tilde G_1^2\tilde G_2}{1+\tilde G_1^2}
 \right]
+\frac{g^4}{16\pi^2}\ 
 \frac{1}{\left[1+\tilde G_1^2 \right]^2}
 \left[
  6-(3+{\alpha^2})\tilde G_1^2\ 
 \right].
\nonumber
\end{eqnarray}
Let us compare above equation with the ladder approximated one,
\begin{eqnarray}
\frac{d \tilde G_2}{dt}
&=&
-2\tilde G_2
+\frac{1}{2\pi^2}\ \frac{\tilde G_2^2}{1+\tilde G_1^2}
 \left[
  1-\frac{2\tilde G_1^2}{1+\tilde G_1^2}
 \right]
\label{e-4fermiL}
\\
&&
+\frac{g^2}{4\pi^2}\ 
 \frac{3+\alpha}{1+\tilde G_1^2}\ 
 \left[
  \tilde G_2-\frac{2\tilde G_1^2\tilde G_2}{1+\tilde G_1^2}
 \right]
+\frac{g^4}{32\pi^2}\frac{(3+\alpha)^2}{\left[1+\tilde G_1^2\right]^2}\ 
 \left[
  1-\tilde G_1^2
 \right].
\nonumber
\end{eqnarray}%
The gauge dependent terms in Eq. (\ref{e-4fermiNL})
should be almost compensated by the gauge dependent anomalous
dimension of the quark fields when we go beyond the LPA,
which will be discussed later.

\section{Numerical Calculation and Results}
\label{sec-Numerical Calculation and Results}

Now we describe how to get the chiral order parameters in QCD 
with our approximation scheme. We work with the Wilsonian
effective potential defined in
Eq. (\ref{e-EffectivePotential-ExtendedExpantion})
with some finite highest
powers $nmax$,  and we numerically integrate its NPRG equation.
The NPRG equation is defined by the $\beta$ function given
in Eq. (\ref{e-GeneralFoumula}),
that is, we take only the quantum loops of the
quarks and gluons and not of the scalar composites.
The initial effective potential is taken from
Eq. (\ref{e-InitialPotential}). 
During evolution the scalar field $\phi$ is fixed to be a certain
value just as an external source field.
The gauge coupling constant is set to follow the one-loop
perturbative $\beta$ function with three flavor quarks.
We take the QCD scale parameter $\LambdaQCD$
to be 490 MeV and we also adopt the same infrared cutoff 
scheme of the gauge coupling constant divergence as in 
Ref. \cite{SDQCD},
since our results should be first compared with the
previous ladder SD results.\cite{SDQCD}\

Integrating the NPRG equation, the effective potential
finally stops to move except for the canonical scaling
behaviors, where the cutoff scale has been lowered well below
the quark mass scale. Then we get the scalar potential
$\tilde G_0(\phi)$ at the fixed $\phi$ value. To solve the
NPRG equation scanning the fixed $\phi$ value, we obtain 
the scalar potential function $\tilde G_0(\phi)$ and its 
minimum point $\langle\phi\rangle$. Then we estimate the chiral
condensates and the quark mass using
Eqs. (\ref{e-condensate}) and (\ref{e-Sigma(0)}). 
The chiral condensates obtained above should be regarded as 
the bare operator condensation at the initial highest cutoff
scale. It should be renormalized through the standard procedure 
to show the renormalized condensates at 1 GeV scale.

We show the results ($\alpha=0$ case) in Fig. 12.
First of all we check the $nmax$ dependence of the results.
Though there still remain some small fluctuations, we may
claim that we have obtained the results of our total subspace
of $nmax = \infty$. We have checked also that the dependence 
of the initial cutoff $\Lambda_0$ should be negligible,
that is, our results are assured to be on the renormalized 
trajectory and we can regard them as those of the infinite initial
cutoff limit.
\begin{figure}[htb]
\begin{minipage}{65mm}
 \epsfxsize=65mm
 \epsfbox{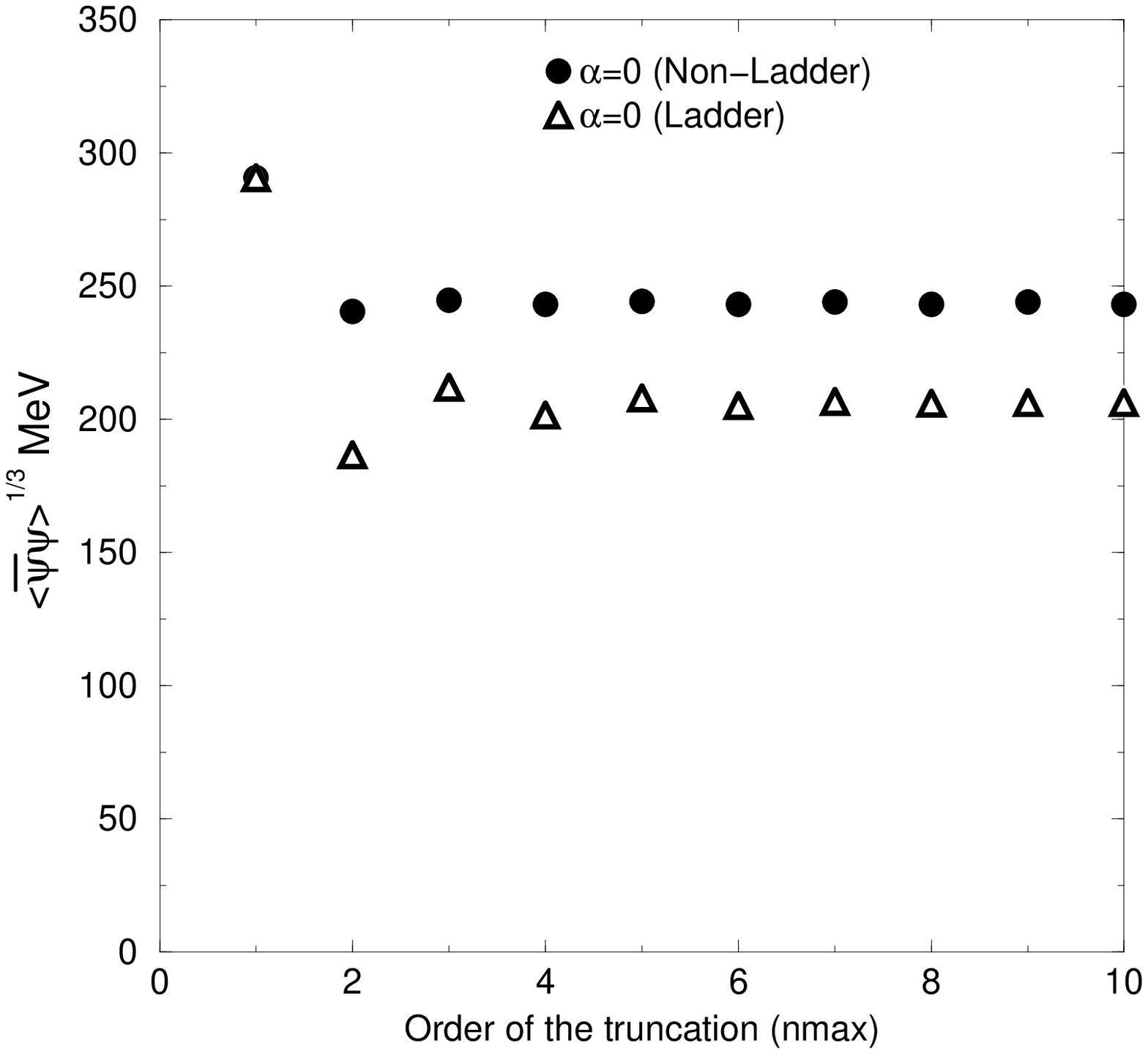} 
\end{minipage}
\hfil
\begin{minipage}{65mm}
 \epsfxsize=65mm
 \epsfbox{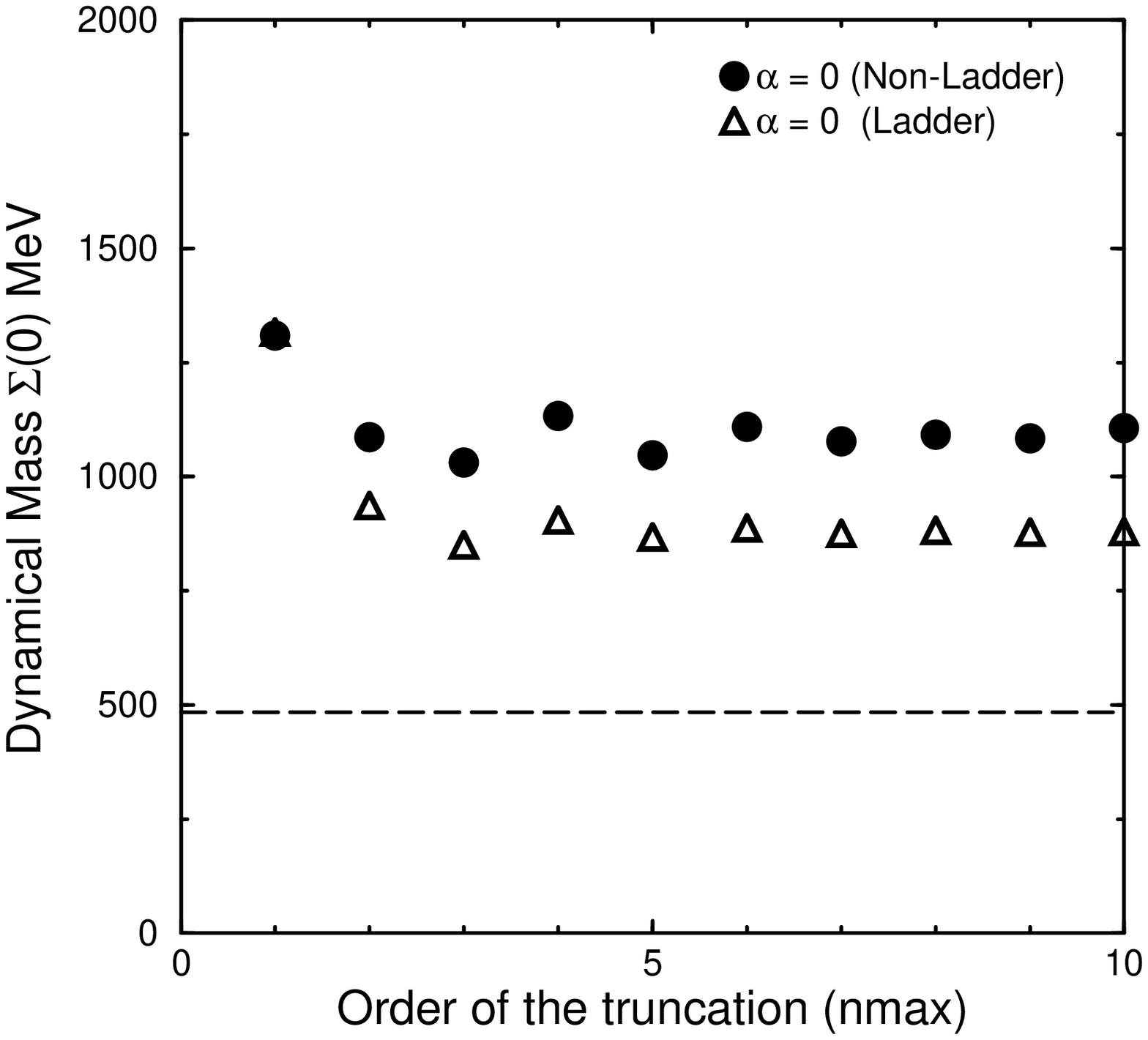} 
\end{minipage}
\centerline{ \parbox{0.9\textwidth}{ 
\caption{\footnotesize 
 The chiral condensates and the dynamical mass of quarks with their
 truncation dependence in the  Landau gauge.  The dashed line is
 $\Lambda_{\rm QCD}$.  Non-ladder results are  enhanced compared to the
 ladder results.   \label{LvsNL}}}}
\hfil
\end{figure}%

Now we compare our non-ladder extended results with the ladder
ones. The ladder results exactly coincide with those of
the ladder SD equation, which assures the total consistency
of our calculational machinery.
The chiral condensates and the quark mass are both
enhanced by including the non-ladder contributions.
Though we do not argue here in detail about the phenomenological
implications of this enhancement, this enhancement is actually 
favorable for phenomenology since our setting of the QCD scale parameter
has been recognized to be much higher than the current estimate
even considering that it is the one-loop $\beta$ estimates.

\begin{figure}[htb]
\begin{minipage}{65mm}
 \epsfxsize=65mm
 \epsfbox{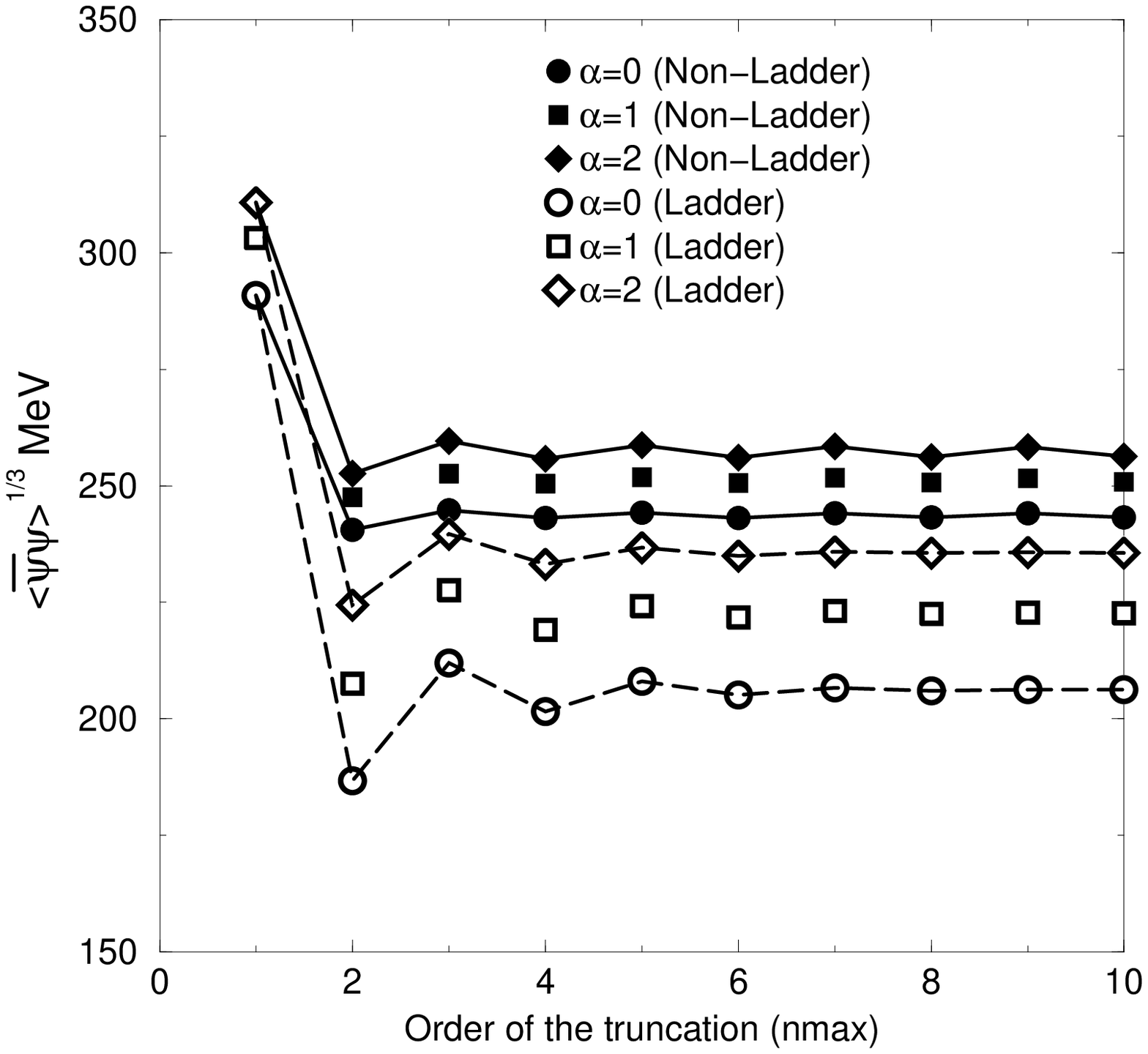}
\end{minipage}
\hfil
\begin{minipage}{65mm}
 \epsfxsize=65mm
 \epsfbox{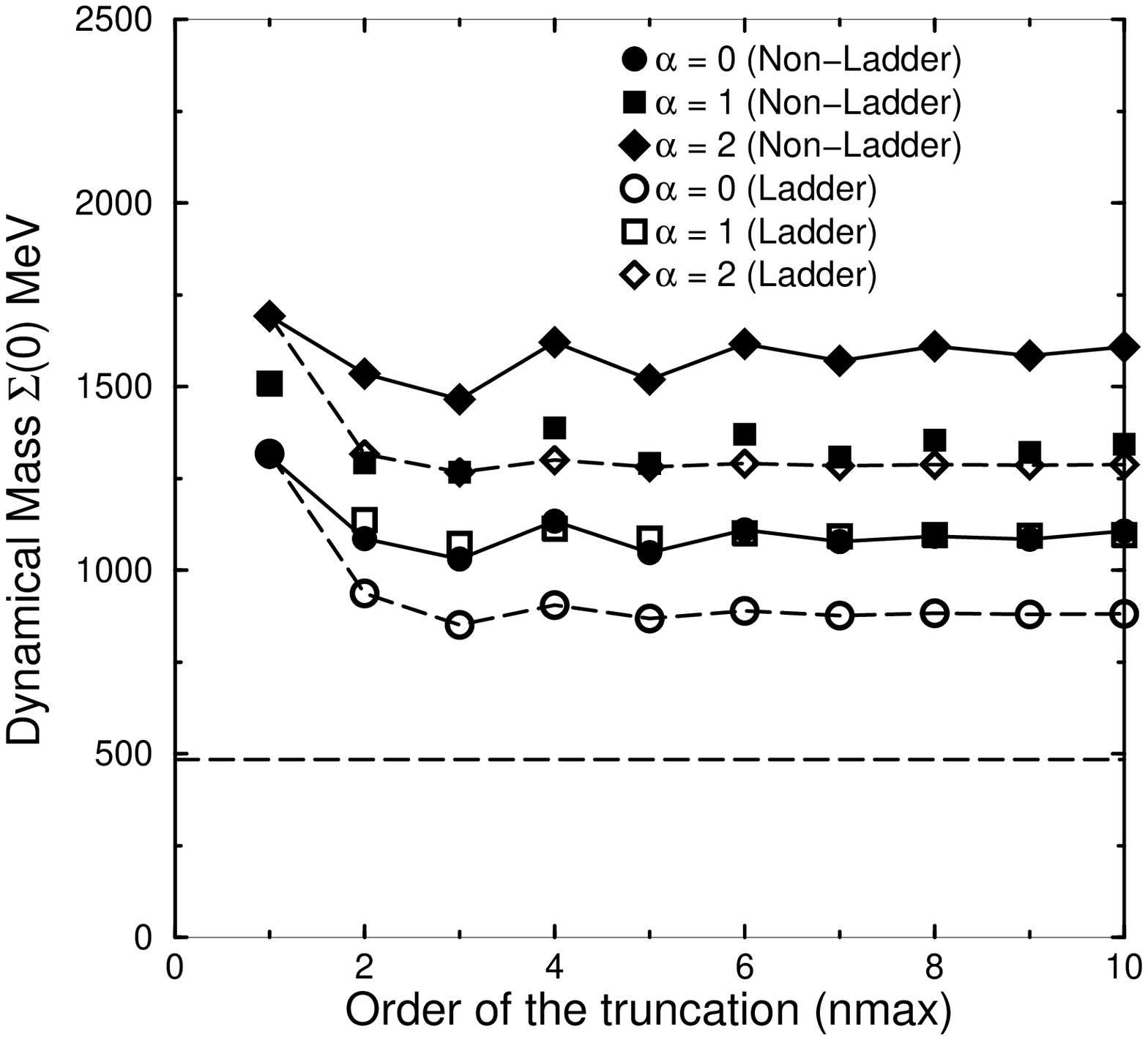} 
\end{minipage}
\centerline{ \parbox{0.9\textwidth}{ 
\caption{\footnotesize 
 The gauge parameter dependence of the chiral condensates and
 the dynamical mass of quarks are plotted for cases $\alpha =0$ (Landau
 gauge), $\alpha =1$ (Feynman gauge) and $\alpha =2$.  
 \label{GaugeDep}}}}
\end{figure}

The gauge parameter dependence of the results are depicted
in Fig. \ref{GaugeDep}.  
Compared with the ladder results, the 
improvement of gauge dependence of $\langle\bar\psi\psi\rangle$ 
is clearly seen in the left figure of Fig. \ref{GaugeDep}.  
On the other hand
the right figure of Fig. \ref{GaugeDep} shows that 
the gauge dependence of $\Sigma(0)$ still remains a lot 
even in the non-ladder approximation. 
We understand the different situations
between these quantities as follows.  
The chiral condensate $\langle\bar\psi\psi\rangle$ is a measurable
physical quantity. 
And it should not depend on the gauge.  
On the other hand the dynamical ``mass'' of quark $\Sigma(0)$ is an
off-shell quantity 
and is not directly related to a measurable quantity; therefore, it may
depend on the gauge.  
It should be noted also that the quark mass $\Sigma(0)$ strongly depends 
on the infrared cutoff scheme of the gauge coupling constant divergence
while the chiral condensates do not.

\section{Issues of the gauge dependence}
\label{sec-Issues of the gauge dependence}
In this section we discuss the origin of the gauge dependence 
in the NPRG method. 
First of all, the Wilsonian effective action $\Seff$ itself
depends on the gauge, 
because it is not directly related to any measurable quantities. 
Therefore the NPRG equations (or the $\beta$ function)
describing the evolution of
the Wilsonian effective action also depend on the gauge.  
Furthermore even at the infrared limit, the effective action is not
totally gauge independent except for the on-shell quantities. 
For example the effective potential is not gauge independent except for
the position of the minimum. 
Thus in general the $\beta$ function which depends on the gauge
parameter finally gives the gauge independent results only for physical
quantities at the infrared limit.
Actually in our approximation scheme of evaluating the effective
potential, there is no way of erasing all the gauge parameter dependence 
in the $\beta$ function.

We discuss here the gauge dependence due to the approximation
we adopted, that is, the Local Potential Approximation.
It ignores any corrections to the derivative couplings including
the kinetic terms, and therefore no anomalous dimension is taken
into account. This seems to be the largest source of the gauge dependence.
We will report elsewhere the results taking account of the quark
anomalous dimension, where we will see the reduced gauge dependence.
Before getting these new
results, we may evaluate the physical quantities as follows.
In the one-loop approximation the quark anomalous dimension 
is proportional to $\alpha$, and therefore
we have vanishing anomalous dimension in the Landau gauge $\alpha=0$.
Therefore we may claim that the Landau gauge results in our scheme
are most significant and they would be very near to the coming results
with quark anomalous dimension.
Then our main results should read,
\begin{eqnarray}
 \left.
  \frac{\langle\bar\psi\psi\rangle^{1/3}}{\LambdaQCD}
 \right|_{\rm\scriptsize non-ladder}
 = 0.512\pm 0.014,
\end{eqnarray}
which is compared with the previous ladder results,
\begin{eqnarray}
 \left.
  \frac{\langle\bar\psi\psi\rangle^{1/3}}{\LambdaQCD}
 \right|_{\rm\scriptsize ladder}
 =0.439.
\end{eqnarray}

\begin{figure}[ht]
 \centerline{
 \begin{minipage}{50mm}
  \epsfxsize=35mm
 \epsfbox{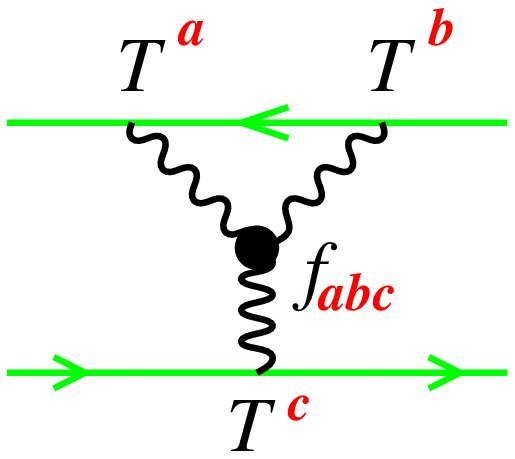}
 \end{minipage}
\hfil
 \begin{minipage}{50mm}
  \epsfxsize=35mm
 \epsfbox{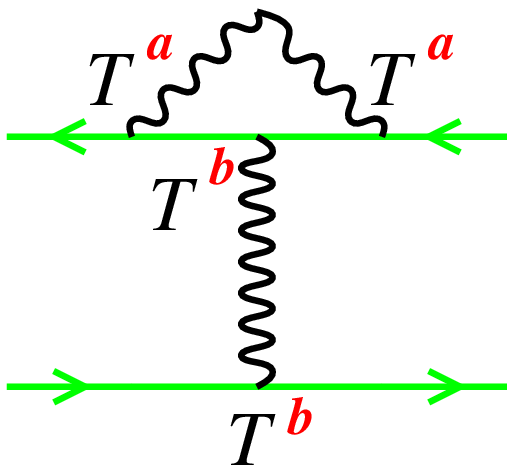}
 \end{minipage}}
\centerline{ \parbox{0.9\textwidth}{ 
 \caption{\footnotesize 
  The ``Penguin''diagrams contributing to the four-fermi $\beta$ function.  
  \label{gamma01}}}}
\end{figure}

\begin{figure}[ht]
\centerline{ \begin{minipage}{50mm}
  \epsfxsize=35mm
 \epsfbox{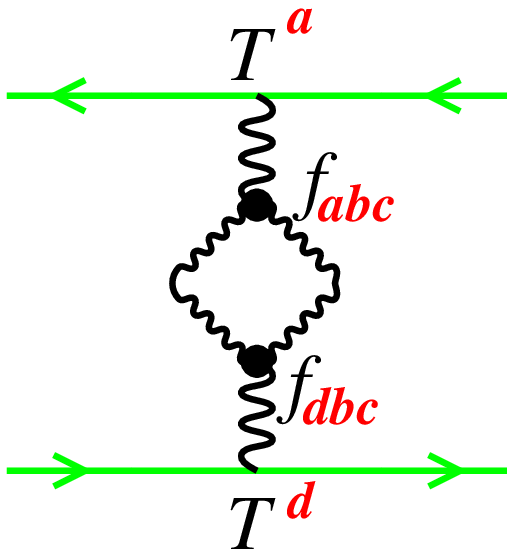}
 \end{minipage}
\hfil
 \begin{minipage}{50mm}
  \epsfxsize=35mm
 \epsfbox{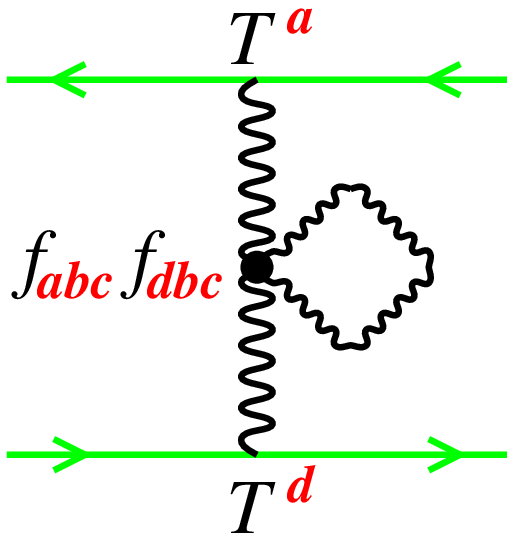}
 \end{minipage}}
\centerline{ \parbox{0.9\textwidth}{ 
 \caption{\footnotesize 
 The diagrams containing the wave function renormalization of
  the gauge field in the four-fermi amplitudes. 
  \label{ZA}}}}
\end{figure}

There are other subtleties of the gauge dependence due to the LPA.
In the above calculation we have done further approximation to ignore
some parts contributing to the $\beta$ functions.
In Eq. (\ref{e-GeneralFoumula}) we omitted the non-Abelian commutator
parts of the
gauge effective vertex. This omission itself does not bring the
gauge dependence. Rather it ``hides'' the gauge dependence of the LPA.
Consider the four-fermi amplitude, for example.
Such commutator parts should be summed
up with diagrams in Fig. \ref{gamma01}
(second derivative part of 
the gluon field, $\partial_\mu F_{\mu\nu} \bar\psi\gamma_\nu\psi$
as an operator form) to generate  
the gauge independent (on-shell) four-fermi amplitudes. 
This situation is quite the same as the Penguin diagram to give the local
four-fermi effective operators.
Of course we also have to add all related diagrams in Fig. \ref{ZA}
and ghost diagrams
to get totally gauge independent results with the properly renormalized
gauge coupling constant.
Therefore we have to take account of the derivative couplings 
$\partial_\mu F_{\mu\nu} \bar\psi\gamma_\nu\psi$
to compensate the gauge dependence appearing in the four-fermi box diagrams.

All these extension requires the higher order derivative 
couplings in our sub-theory space. Then we have to proceed to use
smooth cutoff scheme NPRG equations since the sharp cutoff NPRG
equations suffer singularities when applied to the derivative couplings.

\section{Summary and Discussion}
\label{sec-Summary and Discussion}

In this article we challenge a beyond the ladder calculation of the
dynamical chiral symmetry breaking in QCD by using the non-ladder extension
in the Non-Perturbative Renormalization Group method.
The ladder approximation of the NPRG Local Potential $\beta$ function
has been integrated to give exactly the same results of the (improved) ladder
Schwinger-Dyson equation for the chiral condensates
and the dynamical mass of quark $\Sigma(0)$. 
Extension beyond the ladder has strong
motivation of reducing the inevitable gauge dependence of 
the ladder approximation.

We add non-ladder diagrams to the
NPRG ladder $\beta$ function, trying to reduce the gauge dependence
of the physical results. We develop a set of $\beta$ functions using 
the effective gluon vertex defined by the sum of the ladder and the crossed 
couplings. We numerically solve this new $\beta$ function to get
the chiral condensates and quark mass function at zero momentum.
They are enhanced compared with the previous ladder results, which are
favorable phenomenologically. Also we evaluate the gauge parameter 
dependence of our results and find it is fairly reduced compared to the 
ladder case.

We stress here again that our results are the first results in 
the long history of analyzing the dynamical chiral symmetry breaking in
gauge theories, which goes beyond the (improved) ladder equipping with a
systematic approximation method. 
This is realized by quite a new viewpoint of the NPRG method for the
dynamical chiral symmetry breaking.

\section*{Acknowledgements}
The authors thank to J.-I. Sumi, and K. Morikawa for valuable
discussions.

\end{document}